\documentclass{goeproc}

\def\aap{A\&A}%
\def\apj{ApJ}%
\begin{document}

\title{The height dependence of temperature -- velocity correlation in the solar photosphere}

\author[1,2,*]{J. Koza}
\author[2]{A. Ku\v{c}era}
\author[2]{J. Ryb\'ak}
\author[3]{H. W\"ohl}

\affil[1]{Sterrekundig Instituut, Utrecht University, The Netherlands}
\affil[2]{Astronomical Institute, Slovak Academy of Sciences, Tatransk\'{a} Lomnica, Slovakia}
\affil[3]{ Kiepenheuer-Institut f\"ur Sonnenphysik, Freiburg, Germany}
\affil[*]{\textit{Email:} j.koza@astro.uu.nl}

\runningtitle{Temperature -- velocity correlation in the solar photosphere}
\runningauthor{J.~Koza et al.}

\firstpage{139}

\maketitle

\begin{abstract}
We derive correlation coefficients between temperature and
line-of-sight velocity as a function of optical depth throughout the
solar photosphere for the non-magnetic photosphere and a small area of
enhanced magnetic activity.  The maximum anticorrelation of about
$-0.6$ between temperature and line-of-sight velocity in the
non-magnetic photosphere occurs at $\log\tau_5=-0.4$.  The magnetic
field is another decorrelating factor along with 5-min oscillations
and seeing.
\end{abstract}
\vspace{-1.5mm}
\section{Introduction}
\vspace{-1.5mm}
  The correlative analysis proves to be an essential tool in 
disentangling of causal relations in the solar atmosphere.
  Recently, \citet{RuttenandKrijger2003} and \citet{Ruttenetal2004}
quantified the correlation of the reversed granulation observed in the
low chromosphere  and mid-photosphere  with surface granulation in quest
for the nature of internetwork background brightness patterns in these layers.
  In agreement with these studies \citet{Puschmannetal03} demonstrated
 that filtering out of the p-modes is inevitable for studying the convective
 structures in the solar photosphere because p-modes mostly reduce the correlation
  between various line parameters.
  \citet{Odertetal05} showed that correlation coefficients can fluctuate strongly in time with
amplitudes of over 0.4 due to 5-min oscillations and the amplitudes are larger
for higher formed lines.
  In case of weak lines the situation worsens even more, because correlations
derived from them are influenced stronger by seeing.

  In this paper, we address the dissimilarity
between non-magnetic and magnetic region seen in height variations of
the correlation between temperature and line-of-sight velocity.
  We compare our results 
with a similar study by \citet{RHidalgoetal99}.
  Our analysis follows on the paper
\citet[][henceforth Paper~I]{Kozaetal2006a} and we
invite the reader to have the paper at hand for further references.
\defcitealias{Kozaetal2006a}{Paper~I} 
\begin{figure}[t]
\center{\includegraphics[scale=0.51]{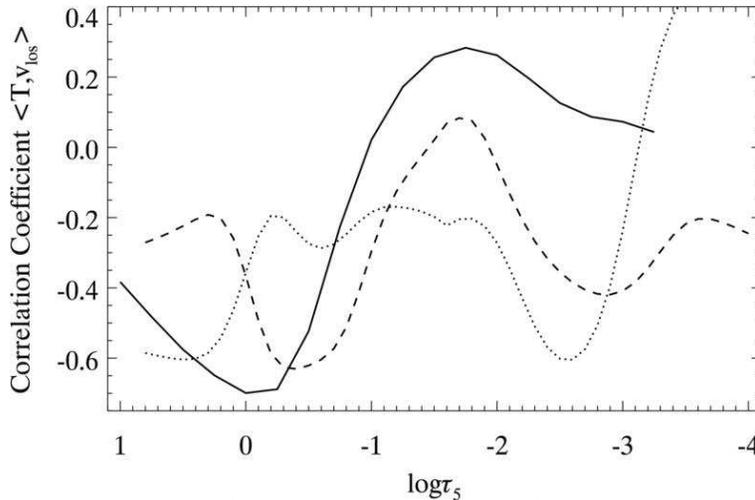}}
\vspace{-0.5cm}
\caption{The height variation of the correlation between line-of-sight velocity and temperature
for the results of \citet{RHidalgoetal99} (solid) and for the non-magnetic (dashed) and magnetic (dotted)
region defined in \citetalias{Kozaetal2006a}.}
\label{koza_fig1}
\end{figure}

\section{Observational data and inversion procedure}
    We use a time sequence of spectrograms obtained at the German Vacuum Tower 
Telescope at the Observatorio del Teide on April 28, 2000. 
    The inversion code SIR \citep{RuizCoboanddelToroIniesta92} was employed
in the analysis of this observation.
Thorough descriptions of the observational data, inversion procedure, and spectral lines
are given in Paper~I.

\section{Results}
\begin{figure}[t]
\center{\includegraphics[scale=0.50]{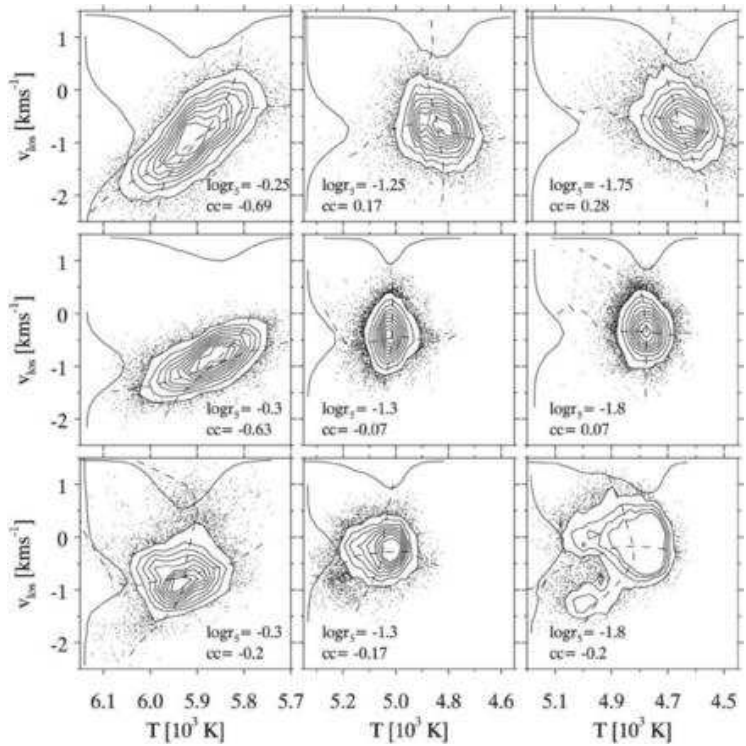}}
\vspace{-0.25cm}
\caption{Height-dependent scatter correlations of the line-of-sight velocity versus temperature.
{\em Top row:} data from \citet[][see p.\,315, Fig.\,1]{RHidalgoetal99}.
{\em Middle and bottom row:} our data for the non-magnetic and magnetic region \citepalias{Kozaetal2006a}, respectively.
 The optical depths $\log\tau_5$ and correlation coefficients {\em cc} are specified at each panel.
 Negative velocities indicate upflows.
 The rescaled total distributions of temperatures and line-of-sight velocities are shown as solid 
 curves at the top and the left side of each panel, respectively. The dashed curves show the first moments
 of the sample density distributions over temperature and velocity bins.
}
\label{koza_fig2}
\end{figure}
  Figure\,\ref{koza_fig1} shows the height variations of the correlation between 
temperature and line-of-sight velocity for three different sets of data.
  The results of \citet{RHidalgoetal99} indicate the significant
anticorrelation between granules and intergranular lanes reaching the
maximum of about $-0.7$ at $\log\tau_5=-0.2$.  The subsequent weakening of this anticorrelation over the
$\log\tau_5\in\langle-0.2,-1.0\rangle$ range is followed by a rise of
correlation up to 0.28 at the middle photosphere at
$\log\tau_5=-1.75$.   No significant correlation exists in the upper
photosphere.
In the lower layers of the non-magnetic region \citepalias{Kozaetal2006a} the
 anticorrelation decreases to $-0.63$
at $\log\tau_5=-0.4$.  However, in the middle photosphere temperature
and line-of-sight velocity are almost uncorrelated with a local peak
value of 0.08 at $\log\tau_5=-1.7$.  Higher up at $\log\tau_5=-2.9$
the anticorrelation of about $-0.42$ is established again.
  In the sub-photospheric layers of the magnetic region the
anticorrelation of $-0.6$ was found at $\log\tau_5=0.5$.
An approximately constant value of anticorrelation $-0.2$
was obtained in the low and middle photosphere.
 In the upper photosphere the anticorrelation reaches again $-0.6$.
   
   Figure\,\ref{koza_fig2} compares temperatures and line-of-sight
velocities in the form  of scatter correlation plots.
Each plotted sample represents temperature and line-of-sight velocity
specified along the {\em x} and {\em y} axes at a given pixel  along
the slit at a time within the interval of 15 min.  From the top down,
the row panels show correlations of the results of
\citet{RHidalgoetal99} and our results
  in the non-magnetic and magnetic region in
three selected optical depths $\log\tau_5=-0.3, -1.3,$ and $-1.8$.  
Plot saturation is avoided by showing
sample density contours rather than individual points, except for the
extreme outliers.  The total distributions of temperatures and
line-of-sight velocities are shown at the top and the left side of
each panel, respectively.
    The first-moment curves are aligned at large correlation and
become perpendicular in the  absence of any correlation
 \citep{RuttenandKrijger2003}.
  The first column in Fig.\,\ref{koza_fig2} shows
good agreement of correlation coefficients and positions of maxima of
velocity distributions in the non-magnetic
region with the results of \citet{RHidalgoetal99}.  However, the
temperature distributions are dissimilar both in terms of asymmetry
and also the positions of maxima.  Our results indicate predominance of
higher temperatures in the sample in contrast with lower temperatures
prevailing in the results of \citet{RHidalgoetal99}. 
In the magnetic region, weak anticorrelation was found.
The temperature distribution in this
region is almost symmetric with maximum at higher temperatures than in
the non-magnetic region.  The second column of Fig.\,\ref{koza_fig2}
corresponds to the layers where granulation is almost erased.
While the temperature distributions in the non-magnetic region  and in
the results of \citet{RHidalgoetal99} are symmetric, in the magnetic
region the asymmetry indicates the abundant higher temperatures.
  The positive correlation in the results of
\citet{RHidalgoetal99} shown in the third column suggests reversed
granulation.  However, this is not seen in our results.  In the
magnetic region the asymmetries of temperature and velocity
distributions indicate higher abundance of relatively hotter pixels
with faster upflows.

\section{Discussion}
Figures\,\ref{koza_fig1} and \ref{koza_fig2} show dissimilarities both in height
variations of correlation and
distributions, although we and \citet{RHidalgoetal99} used
  VTT observations and the SIR code.
Because the maximum of anticorrelation found at sub-photospheric layers
 of the magnetic region is out of the range of sensitivity
 of the spectral lines \citepalias{Kozaetal2006a}, we
 disregard this feature. Very low
 anticorrelation found over $\log\tau_5\in\langle 0.0,-2.0 \rangle$ in
 the magnetic region (Fig.\,\ref{koza_fig1}) suggests that magnetic field
 is another important decorrelating  factor along with 5-min
 oscillations and seeing \citep{Puschmannetal03,Odertetal05}.  In our
 results, the middle layers of the non-magnetic and magnetic region
 lack  signatures of  reversed granulations (Fig.\,\ref{koza_fig1}).  The
 sinusoidal shape of the correlation coefficient in the non-magnetic
 region over the $\log\tau_5\in\langle -1.2, -3.5 \rangle$ range can
 be explained as a sum of  positive correlation
 typical for reversed granulation
 and negative  anticorrelation
 characteristic for 5-min oscillations.

\section{Summary}
   Using a time sequence of high-resolution spectrograms and the SIR inversion code we have
inferred height variation of correlation between the temperature and line-of-sight velocity 
 throughout the quiet solar photosphere and a small magnetic region.
   The most important aspect is comparison of the results with the akin study
 by \citet{RHidalgoetal99}. 
   We found in agreement with \citet{RHidalgoetal99} that the maximum anticorrelation $-0.6$
between the temperature and line-of-sight velocity in the non-magnetic region occurs
at $\log\tau_5=-0.4$.
   The absence of signatures of reversed granulation in the middle layers of
the non-magnetic region is likely to be due to 5-min oscillations, which
 negative anticorrelation prevails over weaker positive correlation
typical for reversed granulation.    
   Our  results show that magnetic field is another decorrelating factor
along with 5-min oscillations and seeing.

\begin{acknowledgements}
  The VTT is operated by the Kiepenheuer-Institut f\"{u}r
Sonnenphysik, Freiburg, at the Observatorio del Teide of the Instituto
de Astrof\'{i}sica de Canarias.  We are grateful to B.~Ruiz Cobo (IAC)
for kindly providing of the original data used in Figs.\,\ref{koza_fig1}
and \ref{koza_fig2}.  This research is part of the European Solar Magnetism Network 
(EC/RTN contract HPRN-CT\discretionary{-}{-}{-}2002-00313). 
This work was supported by the Slovak grant agency
VEGA (2/6195/26) and by the Deut\-sche For\-schungs\-ge\-mein\-schaft
grant (DFG 436 SLK 113/7).   J.~Koza's research is supported by a
Marie Curie  Intra-European Fellowships within the 6th European
Community Framework Programme.
\end{acknowledgements}


\begin{thebibliography}{8}
\expandafter\ifx\csname natexlab\endcsname\relax\def\natexlab#1{#1}\fi

\bibitem[{{Koza} {et~al.}(2006){Koza}, {Ku{\v c}era}, {Ryb{\'a}k}, \&
  {W{\"o}hl}}]{Kozaetal2006a}
{Koza}, J., {Ku{\v c}era}, A., {Ryb{\'a}k}, J., \& {W{\"o}hl}, H. 2006, \aap,
  458, 941, (Paper~I)

\bibitem[{{Odert} {et~al.}(2005){Odert}, {Hanslmeier}, {Ryb{\'a}k}, {Ku{\v
  c}era}, \& {W{\"o}hl}}]{Odertetal05}
{Odert}, P., {Hanslmeier}, A., {Ryb{\'a}k}, J., {Ku{\v c}era}, A., \&
  {W{\"o}hl}, H. 2005, \aap, 444, 257

\bibitem[{{Puschmann} {et~al.}(2003){Puschmann}, {V{\'a}zquez},
  {Bonet}, {Ruiz Cobo}, \& {Hanslmeier}}]{Puschmannetal03}
{Puschmann}, K., {V{\'a}zquez}, M., {Bonet}, J.~A., {Ruiz Cobo}, B., \&
  {Hanslmeier}, A. 2003, \aap, 408, 363

\bibitem[{Rodr\'{\i}guez~Hidalgo {et~al.}(1999)Rodr\'{\i}guez~Hidalgo, {Ruiz
  Cobo}, {Collados}, \& {del Toro Iniesta}}]{RHidalgoetal99}
Rodr\'{\i}guez~Hidalgo, I., {Ruiz Cobo}, B., {Collados}, M., \& {del Toro
  Iniesta}, J.~C. 1999, in ASP Conf. Ser. 173: Stellar Structure: Theory and
  Test of Connective Energy Transport, ed. A.~Gim\'{e}nez, E.~F. Guinan, \&
  B.~Montesinos, 313

\bibitem[{{Ruiz Cobo} \& {del Toro
  Iniesta}(1992)}]{RuizCoboanddelToroIniesta92}
{Ruiz Cobo}, B. \& {del Toro Iniesta}, J.~C. 1992, \apj, 398, 375

\bibitem[{{Rutten} {et~al.}(2004){Rutten}, {de Wijn}, \&
  {S{\"u}tterlin}}]{Ruttenetal2004}
{Rutten}, R.~J., {de Wijn}, A.~G., \& {S{\"u}tterlin}, P. 2004, \aap, 416, 333

\bibitem[{{Rutten} \& {Krijger}(2003)}]{RuttenandKrijger2003}
{Rutten}, R.~J. \& {Krijger}, J.~M. 2003, \aap, 407, 735

\end{thebibliography}
\end{document}